\documentclass[aps,prd,amsfonts,amssymb, amsmath, showpacs, showkeys, a4paper, nofootinbib, superscriptaddress, 11pt, raggedbottom]{revtex4-2}

\usepackage{braket}
\usepackage{bm}
\usepackage{graphicx}
\usepackage{slashed}
\usepackage{subfig}
\usepackage{mathrsfs}
\usepackage{color}
\usepackage{mathtools}
\usepackage[normalem]{ulem}

\begin{document}

\title{\Large Green's function analysis of the Neutron Lloyd interferometer}

\author{Christian K\"{a}ding}
\email{christian.kaeding@tuwien.ac.at}
\affiliation{Technische Universit\"at Wien, Atominstitut, Stadionallee 2, 1020 Vienna, Austria}

\author{Mario Pitschmann}
\email{mario.pitschmann@tuwien.ac.at}
\affiliation{Technische Universit\"at Wien, Atominstitut, Stadionallee 2, 1020 Vienna, Austria}

\author{Hartmut Abele}
\email{hartmut.abele@tuwien.ac.at}
\affiliation{Technische Universit\"at Wien, Atominstitut, Stadionallee 2, 1020 Vienna, Austria}

\begin{abstract}
The neutron optical Lloyd interferometer can serve as a potent experiment for probing fundamental physics beyond the standard models of particles and cosmology. In this article, we provide a full Green's function analysis of a Lloyd interferometer in the limit that the reflecting mirror extends to the screen. We consider two distinct situations: first, we will review the theoretical case of no external fields being present. Subsequently, we will analyze the case in which a gravitational field is acting on the neutrons. The latter case provides the theory necessary for using a Lloyd interferometer as a probe of gravitational fields. 
\end{abstract}


\maketitle


\section{Introduction}
\begin{figure}[htbp]
\begin{center}
\includegraphics[scale=0.7]{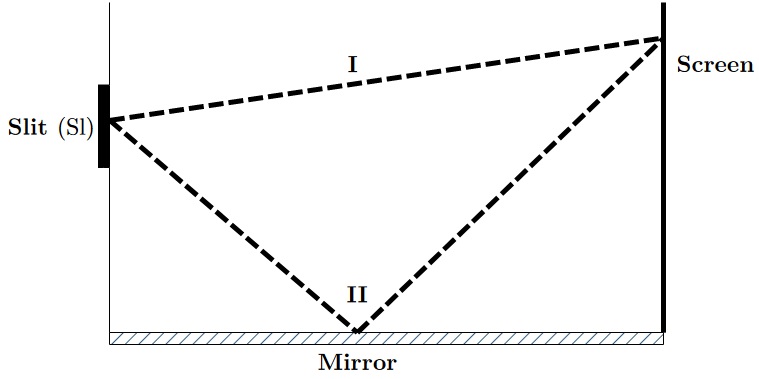}
\caption{Setup of a neutron Lloyd inteferometer; the neutron wave function enters the experiment through a slit $(\mathrm{Sl})$, one path $(\mathrm{I})$ traverses directly to the detection screen, while another path $(\mathrm{II})$ gets reflected at a mirror and subsequently interferes with $(\mathrm{I})$ on the screen.}
\label{Fig:Interfero}
\end{center}
\end{figure}

In 1831 Lloyd \cite{Lloyd} introduced an interferometry experiment, in which two light beams originating from the same slit source interfer with each other after one of them has been reflected on a mirror and the other one propagated directly to the target screen, see Fig.\ref{Fig:Interfero}. Depending on the differences in distances travelled by both beams, an interference pattern can be observed. Since its invention, the optical Lloyd interferometer has seen ample applications, for example, Refs.\,\cite{Ap1967Langenbeck, Ap1970Langenbeck, Ap1970Watkins, Ap1992Kielkopf, Ap1999Rocca, Ap2010Abdullina, Ap2011Wathuthanthri, Ap2013Xinghui, Ap2016Ren, Ap2018Li, Ap2022Rani}. Furthermore, in more recent years, it has been suggested to perform Lloyd interferometry with neutrons instead of light \cite{Gudkov1993, Pokotilovski2011, Filter}. Such proposals are based on ideas related to neutron interferometry \cite{Rauch1974,  Rauch2015}, which is a well-established class of experiments.
\\
Ultra-cold neutrons and neutron optical experiments are excellent means for probing fundamental interactions and symmetries \cite{Abele:2008zz,Dubbers:2011ns}. Examples include the neutron lifetime~\cite{mampe1989neutron,Arzumanov:2000ma,Serebrov:2007ve,ezhov2009magnetic,Pichlmaier:2010zz,Serebrov:2017bzo,Pattie:2017vsj,Ezhov:2014tna} and other decay parameters like $\beta$-decay correlation coefficients~\cite{UCNA:2008pxo,UCNA:2012fhw,UCNA:2017obv}, measurements of its magnetic moment, quantum mechanical~\cite{osti_6125370,rauch2015neutron} or neutron optical~\cite{frank2006effect,frank2011new} properties,  searches for a charge of the neutron~\cite{Durstberger-Rennhofer:2011ghz} and the electric dipole moment. Besides, the search for a permanent electric dipole moment of the neutron investigates a high-energy scale in particle physics that cannot be reached by accelerators on Earth. The present experimental limit on this quantity is $|d_n|$$ <$1.8$\times$10$^{-26}$\,$\textnormal{e\,cm} \,\,\mathrm{(90
\% \, C.L.)}$~\cite{Abel:2020pzs}.
\\
In addition, ultra-cold neutrons and neutron optical experiments have proven themselves to be powerful tools for probing gravity \cite{Abele:2015uua, Jenke:2019qkw, Pitschmann:2019boa, Sedmik:2019twj, Jenke:2020obe, Suda2021, Ivanov:2021bvk, Muto:2022eok} and physics beyond the standard models of particles and cosmology \cite{Lemmel:2015kwa, Ivanov:2016rfs, Brax:2017hna, Cronenberg:2018qxf, Ivanov:2019ouz, Pitschmann:2020ejb, Sponar:2020gfr,  Brax:2022uyh}. Lloyd's mirror is another promising neutron interferometric setup that has even been  considered as a novel way of discovering or constraining fifth forces and new types of scalar fields \cite{Pokotilovski2012, Pokotilovski2013}. However, analyses such as presented in Refs.\,\cite{Pokotilovski2012, Pokotilovski2013} are strongly approximative since they use simplified geometrical path length differences for determining the phase differences. This cannot be sufficient when discussing gravitational or fifth force-inducing scalar fields since they are known to curve the paths on which the neutrons are propagating. For this reason, inspired by the treatment in Ref.\,\cite{Brukner1997}, we present a more accurate analysis based on Green's functions. Green's functions enable us to fully capture the effects on the neutrons induced by external fields and to compute the resulting neutron wave functions necessary for predicting interference patterns in a Lloyd interferometer.
\\
At first, we will review the hypothetical case of no external fields being present in the limit that the reflecting mirror extends to the screen, as it was already discussed by H.~Filter (and M.~Pitschmann) in Ref.\,\cite{Filter}. Afterwards, we extend the discussion by including an external gravitational field. This analysis then provides the theoretical foundation for using a Lloyd interferometer as a probe of gravitational fields. 


\section{Green's functions}

Green's functions are powerful tools for solving inhomogeneous linear differential equations. Before demonstrating how they can be applied in the context of Lloyd interferometry, we will now shortly review how to solve the general Schr\"odinger equation for a particle of mass $m$ using them.
\\
We start by making a stationary wave approximation for the wave function
\begin{eqnarray}\label{Eq:StatWave}
\Psi(\mathbf{r},t) &=& \psi(\mathbf{r})e^{-\mathrm{i}\omega t}\,\,\,
\end{eqnarray}
which we then substitute into the Schr\"odinger equation:
\begin{eqnarray}\label{Eq:Schroedinger}
\left( -\frac{\hbar^2}{2m} \Delta + V(\mathbf{r}) \right)\Psi(\mathbf{r},t) &=& \mathrm{i}\hbar \frac{\partial}{\partial t} \Psi(\mathbf{r},t)\,\,\,.
\end{eqnarray}
This leads us to the Helmholtz equation
\begin{eqnarray}\label{Eq:Helmholtz}
\left( \Delta + k^2 \right)\psi(\mathbf{r}) &=& 0\,\,\,,
\end{eqnarray}
where $k^2 = \frac{2m}{\hbar^2}(\hbar\omega - V(\mathbf{r}))$. The inhomogeneous Helmholtz equation may be solved via a Green's function $G(\mathbf{r},\mathbf{r'})$ fulfilling
\begin{eqnarray}\label{Eq:GreensEqn}
\left( \Delta + k^2 \right)G(\mathbf{r},\mathbf{r'}) &=& -4\pi \delta^{(3)}(\mathbf{r}-\mathbf{r'})\,\,\,,
\end{eqnarray}
which gives
\begin{eqnarray}\label{Eq:GreensFunc}
G(\mathbf{r},\mathbf{r'}) &=& \frac{e^{\mathrm{i}k|\mathbf{r}-\mathbf{r'}|}}{|\mathbf{r}-\mathbf{r'}|}\,\,\,.
\end{eqnarray}
Using Eqs.\,(\ref{Eq:Helmholtz}) and (\ref{Eq:GreensEqn}), we can easily show that for $\mathbf{r} \in V $ the following holds:
\begin{eqnarray}\label{Eq:Green1}
\int_V d^3r' \left\{ \psi(\mathbf{r'}) (\Delta' + k^2) G(\mathbf{r},\mathbf{r'}) - G(\mathbf{r},\mathbf{r'})(\Delta' + k^2)\psi(\mathbf{r'})  \right\} &=& -4\pi \psi(\mathbf{r})  \,\,\,.
\end{eqnarray}
In addition, employing Green's theorem, we know 
\begin{eqnarray}\label{Eq:Green2}
\int_V d^3r' \left\{ \psi(\mathbf{r'}) (\Delta' + k^2) G(\mathbf{r},\mathbf{r'}) \right.
\,\,\,\,\,\,\,\,
&\phantom{=}&
\nonumber
\\
\left. - G(\mathbf{r},\mathbf{r'})(\Delta' + k^2)\psi(\mathbf{r'})  \right\} &=& \oint_{\partial V} d \mathbf{S'} \left\{ \psi(\mathbf{r'}) \nabla' G(\mathbf{r},\mathbf{r'}) - G(\mathbf{r},\mathbf{r'})\nabla' \psi(\mathbf{r'})  \right\}\,\,\,,
\end{eqnarray}
where $d \mathbf{S'}$ is pointing outwards. Combining Eqs.\,(\ref{Eq:Green1}) and (\ref{Eq:Green2}) finally solves the Helmholtz equation (\ref{Eq:Helmholtz}) in terms of the Green's function from Eq.\,(\ref{Eq:GreensFunc}):
\begin{eqnarray}\label{Eq:GenSol}
\psi(\mathbf{r}) &=& \frac{1}{4\pi}\oint_{\partial V} d \mathbf{S'} \left\{ \psi(\mathbf{r'}) \nabla' G(\mathbf{r},\mathbf{r'}) - G(\mathbf{r},\mathbf{r'})\nabla' \psi(\mathbf{r'})  \right\}\,\,\,.
\end{eqnarray}


\subsection{No external fields}

\begin{figure}[htbp]
\begin{center}
\includegraphics[scale=0.7]{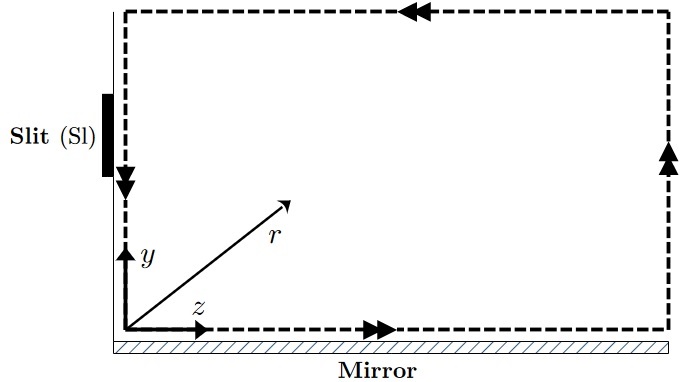}
\caption{Surface integration for a Lloyd interferometer with corner geometry; the entrance slit is reaching from $(-\delta/2,0)$ to $(+\delta/2,0)$ in the $yz$-plane, and the double arrows follow the dotted closed integration path. }
\label{Fig:LloydGreen}
\end{center}
\end{figure}

Finally, we will focus on applications to Lloyd's interferometer. At first, we will look at the idealistic case of no external field being present and compute the corresponding Green's function, which we then use to solve for the neutron wave function. We will focus on the setup of a Lloyd interferometer as presented in Fig.\,\ref{Fig:LloydGreen}.

\subsubsection{Exact solution}
By the method of mirror charges, which is well-known from electrostatics \cite{SchwingerEDyn}, we obtain a Green's function
\begin{eqnarray}\label{Eq:GreenFuncLloyd}
G(\mathbf{r},\mathbf{r'}) &=& 
\sum\limits_{a,b=\pm} ab \frac{ e^{\mathrm{i}k\sqrt{(x-x')^2+(y-ay')^2+(z-bz')^2}} }{\sqrt{(x-x')^2+(y-ay')^2+(z-bz')^2}}
\,\,\,,
\end{eqnarray}
which vanishes everywhere along the dotted integration path given in Fig.\,\ref{Fig:LloydGreen}. We introduced $\mathbf{r} = (x,y,z)^T$ and $\mathbf{r}' = (x',y',z')^T$. In addition, the wave function $\psi(\mathbf{r'})$ also vanishes everywhere on this path except for the part coinciding with the entrance slit. There we have $\psi(\mathbf{r'}) = C e^{\mathrm{i}kz}$ with $C$ being some normalization constant. Using these properties together with the general solution in Eq.\,(\ref{Eq:GenSol}), we find
\begin{eqnarray}\label{Eq:PsiLloyd}
\psi(\mathbf{r}) &=& - \frac{1}{4\pi}\int_{\mathrm{Sl}} d\mathbf{S'} \psi(\mathbf{r'}) \nabla' G(\mathbf{r},\mathbf{r'}) 
\nonumber
\\
&=& \frac{1}{4\pi} \int_{-\infty}^\infty dx' \int^{y_{\mathrm{Sl}} +\delta/2 }_{y_{\mathrm{Sl}} - \delta/2 }  dy' \psi(\mathbf{r'}) \frac{\partial}{\partial z'} G(\mathbf{r},\mathbf{r'})\,\,\,, 
\end{eqnarray}
where $\mathrm{Sl}$ denotes the entrance slit area, and $y_{\mathrm{Sl}}$ and $\delta$ are the slit's center and length on the $y$-coordinate. Substituting Eq.\,(\ref{Eq:GreenFuncLloyd}) into Eq.\,(\ref{Eq:PsiLloyd}), we find
\begin{eqnarray}
\psi(\mathbf{r}) &=& \frac{C}{2\pi} \frac{\partial}{\partial z} \int^{y_{\mathrm{Sl}} +\delta/2 }_{y_{\mathrm{Sl}} - \delta/2 }  dy'  \int_{-\infty}^\infty dx' \left[\frac{ e^{\mathrm{i}k\sqrt{(x-x')^2+(y+y')^2+z^2}} }{\sqrt{(x-x')^2+(y+y')^2+z^2}}
-
\frac{ e^{\mathrm{i}k\sqrt{(x-x')^2+(y-y')^2+z^2}} }{\sqrt{(x-x')^2+(y-y')^2+z^2}} \right]
\nonumber
\\
&=&
\frac{C}{2\pi} \frac{\partial}{\partial z} \int^{y_{\mathrm{Sl}} +\delta/2 }_{y_{\mathrm{Sl}} - \delta/2 }  dy'  \int_{-\infty}^\infty dx \left[ \frac{ e^{\mathrm{i}k\sqrt{x^2+(y+y')^2+z^2}} }{\sqrt{x^2+(y+y')^2+z^2}}
-
\frac{ e^{\mathrm{i}k\sqrt{x^2+(y-y')^2+z^2}} }{\sqrt{x^2+(y-y')^2+z^2}} \right]
\,\,\,.
\end{eqnarray}
Next, we evaluate the $x$-integral via the following integral representation of the Hankel function of the first kind \cite{SchwingerEDyn}: 
\begin{eqnarray}
\mathrm{i}\pi H_0^{(1)}(k\rho) &=& \int^\infty_{-\infty} dz \frac{e^{\mathrm{i}k\sqrt{\rho^2 + z^2}}}{\sqrt{\rho^2 + z^2}} \,\,\,.
\end{eqnarray}
In this way, we obtain
\begin{eqnarray}
\psi(\mathbf{r}) &=& \frac{\mathrm{i} C}{2} \frac{\partial}{\partial z} \int^{y_{\mathrm{Sl}} +\delta/2 }_{y_{\mathrm{Sl}} - \delta/2 }  dy' \left[ H_0^{(1)}(k\sqrt{(y+y')^2 + z^2}) - H_0^{(1)}(k\sqrt{(y-y')^2 + z^2}) \right] \,\,\,.
\end{eqnarray}
Next, using \cite{Watson}
\begin{eqnarray}\label{Eq:HankelAbl}
\frac{d}{dz} H_n^{(1)} (z) &=& \frac{nH_n^{(1)} (z)}{z} - H_{n+1}^{(1)} (z) \,\,\,,
\end{eqnarray}
we find
\begin{eqnarray}
\psi(\mathbf{r}) &=& C \frac{\mathrm{i}  kz}{2}  \int^{y_{\mathrm{Sl}} +\delta/2 }_{y_{\mathrm{Sl}} - \delta/2 }  dy' \left[ \frac{H_1^{(1)}(k\sqrt{(y-y')^2 + z^2})}{\sqrt{(y-y')^2 + z^2}} - \frac{H_1^{(1)}(k\sqrt{(y+y')^2 + z^2})}{\sqrt{(y+y')^2 + z^2}}  \right] \,\,\,,
\end{eqnarray}
and ultimately
\begin{eqnarray}\label{Eq:PsiHank}
\psi(\mathbf{r}) &=& C \frac{\mathrm{i}  kz}{2}   \left[ \int^{y_{\mathrm{Sl}} +\delta/2 - y}_{y_{\mathrm{Sl}} - \delta/2 -y}  dy' \frac{H_1^{(1)}(k\sqrt{y'^2 + z^2})}{\sqrt{y'^2 + z^2}} -
\int^{y_{\mathrm{Sl}} +\delta/2 +y}_{y_{\mathrm{Sl}} - \delta/2 +y}  dy'
\frac{H_1^{(1)}(k\sqrt{y'^2 + z^2})}{\sqrt{y'^2 + z^2}}  \right] \,\,\,.
\end{eqnarray}
\subsubsection{Asymptotic expression}
We will now find an approximation for Eq.\,(\ref{Eq:PsiHank}) by approximating its integrals in the asymptotic case that $z$ is very large and $\delta \to 0$. For this, we first use the asymptotic expression \cite{SchwingerEDyn}
\begin{eqnarray}
H_1^{(1)}(z) &\to& \sqrt{\frac{2}{\pi z}} e^{\mathrm{i} (z-3\pi/4)}\,\,\,,
\end{eqnarray}
which allows us to write Eq.\,(\ref{Eq:PsiHank}) in the asymptotic case of $z$ being large as
\begin{eqnarray}\label{Eq:PsiLargez}
\psi(\mathbf{r}) &\to& C z \sqrt{\frac{k}{2\pi}}  
e^{-\mathrm{i}\pi/4}
\left[ \int^{y_{\mathrm{Sl}} +\delta/2 - y}_{y_{\mathrm{Sl}} - \delta/2 -y}  dy' \frac{e^{\mathrm{i}k\sqrt{y'^2+z^2}}}{(y'^2+z^2)^{3/4}} -
\int^{y_{\mathrm{Sl}} +\delta/2 +y}_{y_{\mathrm{Sl}} - \delta/2 +y}  dy'
\frac{e^{\mathrm{i}k\sqrt{y'^2+z^2}}}{(y'^2+z^2)^{3/4}}  \right] \,\,\,.
\end{eqnarray}
Next, we use that for a convergent integrand and a small integration interval the following approximation holds:
\begin{eqnarray}
\int^{A+\delta/2}_{A-\delta/2} dy f(y) &=& \int^{+\delta/2}_{-\delta/2} dy f(y+A) \,=\, \frac{\delta}{2}\int^{+1}_{-1} dy  f\left( \frac{\delta}{2}y + A \right) \,\to\, \delta f(A)\,\,\,.
\end{eqnarray}
Applying this to Eq.\,(\ref{Eq:PsiLargez}), we find that the wave function for large $z$ and $\delta \to 0$ can be approximated as
\begin{eqnarray}
\psi(\mathbf{r}) &\to& \delta C z \sqrt{\frac{k}{2\pi}}  
e^{-\mathrm{i}\pi/4}
\left\{ \frac{e^{\mathrm{i}k\sqrt{(y_{\mathrm{Sl}} - y)^2+z^2}}}{[(y_{\mathrm{Sl}} - y)^2+z^2]^{3/4}} -
\frac{e^{\mathrm{i}k\sqrt{(y_{\mathrm{Sl}}  +y)^2+z^2}}}{[(y_{\mathrm{Sl}}  +y)^2+z^2]^{3/4}}  \right\} \,\,\,.
\end{eqnarray}


\subsection{Gravitational field}

Now we will consider a physically realistic situation by introducing an external gravitational field. At first, we will derive a general expression for the Green's function for this particular case, following the treatment in Ref.\,\cite{SchwingerEDyn}. Later, we will apply the result explicitly to Lloyd interferometry. 
\subsubsection{General solution}
We consider a gravitational field in $x$-direction and want to find the Green's function for the Helmholtz equation
\begin{eqnarray}
\mathfrak{L}\psi(\mathbf{r}) &=&0\,\,\,,
\end{eqnarray}
where the field operator is given by
\begin{eqnarray}\label{Eq:FieldOperator}
\mathfrak{L} &:=& \Delta_\perp + \partial_x^2 + \frac{2m}{\hbar^2}(E-mgx)\,\,\,,
\end{eqnarray}
and we use the ansatz
\begin{eqnarray}\label{Eq:Gansatz}
G(\mathbf{r},\mathbf{r'}) &=& 4\pi \int \frac{d^2k_\perp}{(2\pi)^2} e^{\mathrm{i} \mathbf{k}_\perp \cdot ( \mathbf{x}_\perp - \mathbf{x'}_\perp ) } \mathrm{g}(x,x';\mathbf{k}_\perp)
\end{eqnarray}
for the Green's function.
Consequently, acting with the field operator from Eq.\,(\ref{Eq:FieldOperator}) on the Green's function gives us the following condition:
\begin{eqnarray}
\mathfrak{L} G(\mathbf{r},\mathbf{r'}) &=& 4\pi \int \frac{d^2k_\perp}{(2\pi)^2} e^{\mathrm{i} \mathbf{k}_\perp \cdot ( \mathbf{x}_\perp - \mathbf{x'}_\perp ) } \left[ \partial_x^2 - \mathbf{k}_\perp^2 + \frac{2m}{\hbar^2}(E-mgx) \right] \mathrm{g}(x,x';\mathbf{k}_\perp)
\nonumber
\\
&=& 4\pi \int \frac{d^2k_\perp}{(2\pi)^2} e^{\mathrm{i} \mathbf{k}_\perp \cdot ( \mathbf{x}_\perp - \mathbf{x'}_\perp ) } \left[ \partial_x^2  + \frac{2m}{\hbar^2}(\tilde{E}-mgx) \right] \mathrm{g}(x,x';\mathbf{k}_\perp)
\nonumber
\\
&=& -4\pi \delta^{(3)}(\mathbf{r}-\mathbf{r'})
\end{eqnarray}
with $\tilde{E} := E- \hbar^2\mathbf{k}_\perp^2/2m$. Subsequently, we extract 
\begin{eqnarray}\label{Eq:Extract}
\left[ \partial_x^2  + \frac{2m}{\hbar^2}(\tilde{E}-mgx) \right] \mathrm{g}(x,x';\mathbf{k}_\perp) &=& -\delta(x-x')\,\,\,.
\end{eqnarray}
Assuming that $\mathrm{g}$ has support only on $\{x\!: x\in [x' -0,x'+0]\}$, integrating Eq.\,(\ref{Eq:Extract}) over $x$ gives
\begin{eqnarray}\label{Eq:gCond1}
\left. \partial_x \mathrm{g}(x,x';\mathbf{k}_\perp) \right|^{x'+0}_{x'-0} &=& -1 \,\,\,.
\end{eqnarray}
Next, we return to Eq.\,(\ref{Eq:Extract}) and multiply it with $x$ from the left, which leads us to
\begin{eqnarray}
\partial_x\left[ x\partial_x \mathrm{g}(x,x';\mathbf{k}_\perp) - \mathrm{g}(x,x';\mathbf{k}_\perp)  \right] + \frac{2mx}{\hbar^2}(\tilde{E}-mgx)  \mathrm{g}(x,x';\mathbf{k}_\perp) &=& -x'\delta(x-x')\,\,\,.
\end{eqnarray}
Integrating this over $x$, we find
\begin{eqnarray}\label{Eq:gCond2}
\left. x\partial_x \mathrm{g}(x,x';\mathbf{k}_\perp) \right|^{x'+0}_{x'-0} -\left.  \mathrm{g}(x,x';\mathbf{k}_\perp) \right|^{x'+0}_{x'-0} &=& -x' \,\,\,.
\end{eqnarray}
Combining Eqs.\,(\ref{Eq:gCond1}) and (\ref{Eq:gCond2}), gives
\begin{eqnarray}\label{Eq:gCond3}
\left.  \mathrm{g}(x,x';\mathbf{k}_\perp) \right|^{x'+0}_{x'-0} &=& 0
\,\,\,.
\end{eqnarray}
For later convenience, we are now going to prove the reciprocity relation
\begin{eqnarray}\label{Eq:Reciprocity}
\mathrm{g}(x,x';\mathbf{k}_\perp) &=& \mathrm{g}(x',x;\mathbf{k}_\perp)\,\,\,.
\end{eqnarray}
For this, we begin by multiplying Eq.\,(\ref{Eq:Extract}) by $\mathrm{g}(x,x'';\mathbf{k}_\perp)$, and then subtracting a copy of the resulting equation but with $x' \leftrightarrow x''$, which results in
\begin{eqnarray}
-\mathrm{g}(x,x'';\mathbf{k}_\perp)\delta(x-x') 
+ \mathrm{g}(x,x';\mathbf{k}_\perp)\delta(x-x'')  
&=& 
\mathrm{g}(x,x'';\mathbf{k}_\perp) \partial_x^2 \mathrm{g}(x,x';\mathbf{k}_\perp) 
\nonumber
\\
&&\,\,\,\,\,\,\,\,\,\,\,\,\,\,\,\,\,\,\,\,\,\,\,\,\,\,\,\,\,\,\,\,
- \mathrm{g}(x,x';\mathbf{k}_\perp) \partial_x^2 \mathrm{g}(x,x'';\mathbf{k}_\perp)
\nonumber
\\
&=&
\partial_x\left[\mathrm{g}(x,x'';\mathbf{k}_\perp) \partial_x \mathrm{g}(x,x';\mathbf{k}_\perp) \right.
\nonumber
\\
&&\,\,\,\,\,\,\,\,\,\,\,\,\,\,\,\,\,\,\,\,\,\,\,\,\,\,\,\,\,\,\,\,\,\,\,\,\,\,\,\,\,
\left.
- \mathrm{g}(x,x';\mathbf{k}_\perp) \partial_x \mathrm{g}(x,x'';\mathbf{k}_\perp)\right]
\,\,\,.\,\,\,\,\,\,\,\,\,\,\,\,\,\,\,\,\,\,
\end{eqnarray}
Integrating this over all of $x$ yields
\begin{eqnarray}\label{Eq:ProofRecip}
-\mathrm{g}(x',x'';\mathbf{k}_\perp) + \mathrm{g}(x'',x';\mathbf{k}_\perp)  
&=& \left[\mathrm{g}(x,x'';\mathbf{k}_\perp) \partial_x \mathrm{g}(x,x';\mathbf{k}_\perp) - \mathrm{g}(x,x';\mathbf{k}_\perp) \partial_x \mathrm{g}(x,x'';\mathbf{k}_\perp)\right] |^{+\infty}_{-\infty}\,\,\,.\,\,\,\,\,\,\,\,\,\,\,
\end{eqnarray}
Since $\mathrm{g}$ has support only on $\{x\!: x\in [x' -0,x'+0]\}$, we know that the right-hand side of Eq.\,(\ref{Eq:ProofRecip}) must vanish, leaving us with the reciprocity relation
\begin{eqnarray}
\mathrm{g}(x',x'';\mathbf{k}_\perp)  
&=& \mathrm{g}(x'',x';\mathbf{k}_\perp)\,\,\,,
\end{eqnarray}
which concludes the proof.
\\
Now we continue with the computation of the function $\mathrm{g}(x,x';\mathbf{k}_\perp) $. For this, we use that for $x\neq x'$ Eq.\,(\ref{Eq:Extract}) takes on the form
\begin{eqnarray}\label{Eq:gEqxneqxs}
\left[ \partial_x^2  + \frac{2m}{\hbar^2}(\tilde{E}-mgx) \right] \mathrm{g}(x,x';\mathbf{k}_\perp) &=& 0\,\,\,.
\end{eqnarray}
In this way, we can find a solution for the Green's function via the homogeneous solution. Following Ref.\,\cite{Pitschmann:2019boa}, the convergent solution of Eq.\,(\ref{Eq:gEqxneqxs}) for $x\geq x'$ is
\begin{eqnarray}\label{Eq:gSol1}
\mathrm{g}(x,x';\mathbf{k}_\perp) &=& C_1(x') \mathrm{Ai}(\sigma)
\end{eqnarray}
with a dimensionless variable
\begin{eqnarray}
\sigma &:=& \sqrt[3]{\frac{2m^2g}{\hbar^2}}\left( x - \frac{\tilde{E}}{mg} \right)\,\,\,,
\end{eqnarray}
while for $x\leq x'$ it is
\begin{eqnarray}\label{Eq:gSol2}
\mathrm{g}(x,x';\mathbf{k}_\perp) &=& C_2(x') \mathrm{Ai}(\sigma) + C_3(x')\mathrm{Bi}(\sigma) 
\,\,\,.
\end{eqnarray}
$\mathrm{Ai}$ and $\mathrm{Bi}$ are Airy functions of the first and second kind, respectively.
Applying the conditions in Eqs.\,(\ref{Eq:gCond1}) and (\ref{Eq:gCond3}) to the solutions in Eqs.\,(\ref{Eq:gSol1}) and (\ref{Eq:gSol2}) leads us to
\begin{eqnarray}
\label{Eq:CCond1}
&&C_1(x')\mathrm{Ai}'(\sigma') - C_2(x')\mathrm{Ai}'(\sigma') - C_3(x') \mathrm{Bi}'(\sigma') \,=\, -\sqrt[3]{\frac{\hbar^2}{2m^2g}}\,\,\,,
\\
\label{Eq:CCond2}
&&C_1(x')\mathrm{Ai}(\sigma') \,-\, C_2(x')\mathrm{Ai}(\sigma') \,-\, C_3(x') \mathrm{Bi}(\sigma')
\,\,=\, 0\,\,\,,
\end{eqnarray}
where
\begin{eqnarray}
\sigma' &:=& \sqrt[3]{\frac{2m^2g}{\hbar^2}}\left( x' - \frac{\tilde{E}}{mg} \right)\,\,\,.
\end{eqnarray}
Combining Eqs.\,(\ref{Eq:CCond1}) and (\ref{Eq:CCond2}) gives
\begin{eqnarray}
C_1(x') &=& C_2(x')+ \sqrt[3]{\frac{\hbar^2}{2m^2g}}\frac{\mathrm{Bi}(\sigma')}{W(\mathrm{Ai},\mathrm{Bi})(\sigma')}\,\,\,,\,\,\,\,\,\,\,\,\,
C_3(x') \,=\, \sqrt[3]{\frac{\hbar^2}{2m^2g}}\frac{\mathrm{Ai}(\sigma')}{W(\mathrm{Ai},\mathrm{Bi})(\sigma')}\,\,\,,
\end{eqnarray}
where $W(\mathrm{Ai},\mathrm{Bi})(\sigma)$ is the Wronskian of $\mathrm{Ai}$ and $\mathrm{Bi}$. Using the fact that every Airy function $f(\sigma)$ must fulfill $f''(\sigma) = \sigma f(\sigma)$, it is straightforward to show that this Wronskian must be constant:
\begin{eqnarray}
\frac{d}{d\sigma}W(\mathrm{Ai},\mathrm{Bi})(\sigma) &=& \mathrm{Ai}(\sigma)\mathrm{Bi}''(\sigma) - \mathrm{Bi}(\sigma)\mathrm{Ai}''(\sigma)
\nonumber
\\
&=& \sigma\mathrm{Ai}(\sigma)\mathrm{Bi}(\sigma) - \sigma\mathrm{Bi}(\sigma)\mathrm{Ai}(\sigma)
\nonumber
\\
&=& 0\,\,\,,
\end{eqnarray}
which implies $W(\mathrm{Ai},\mathrm{Bi})(\sigma) = W(\mathrm{Ai},\mathrm{Bi})(0)$. Taking the values for the Airy functions and their derivatives at $\sigma = 0$, we can show that 
\begin{eqnarray}
W(\mathrm{Ai},\mathrm{Bi})(\sigma) &=& \frac{1}{\pi}
\end{eqnarray}
holds.
So, substituting this into Eqs.\,(\ref{Eq:gSol1}) and (\ref{Eq:gSol2}) leaves us with
\begin{eqnarray}
\mathrm{g}(x,x';\mathbf{k}_\perp) &=& 2C_2(x')\mathrm{Ai}(\sigma) + 
\sqrt[3]{\frac{\hbar^2\pi^3}{2m^2g}}\left[ \Theta(x-x')\mathrm{Bi}(\sigma')\mathrm{Ai}(\sigma) + \Theta(x'-x) \mathrm{Ai}(\sigma')\mathrm{Bi}(\sigma) \right]
\,\,\,.\,\,\,\,\,\,\,\,\,\,\,\,
\end{eqnarray}
Furthermore, from the reciprocity relation (\ref{Eq:Reciprocity}) we conclude
\begin{eqnarray}
C_2(x') &=& \frac{\lambda}{2} \mathrm{Ai}(\sigma')\,\,\,,
\end{eqnarray}
where $\lambda$ is some constant. In consequence, we have
\begin{eqnarray}\label{Eq:Finalg}
\mathrm{g}(x,x';\mathbf{k}_\perp) &=& \lambda\mathrm{Ai}(\sigma)\mathrm{Ai}(\sigma') + 
\sqrt[3]{\frac{\hbar^2\pi^3}{2m^2g}}\left[ \Theta(x-x')\mathrm{Ai}(\sigma)\mathrm{Bi}(\sigma') + \Theta(x'-x) \mathrm{Ai}(\sigma')\mathrm{Bi}(\sigma) \right]
\,\,\,.\,\,\,\,\,\,\,\,\,\,\,\,
\end{eqnarray}
Substituting Eq.\,(\ref{Eq:Finalg}) into the ansatz in Eq.\,(\ref{Eq:Gansatz}), we obtain the solution for the Green's function as
\begin{eqnarray}\label{Eq:GFinalSol}
G(\mathbf{r},\mathbf{r'}) &=& 4\pi \int \frac{d^2k_\perp}{(2\pi)^2} e^{\mathrm{i} \mathbf{k}_\perp \cdot ( \mathbf{x}_\perp - \mathbf{x'}_\perp ) } 
\nonumber
\\
&&
\times
\left\{ \lambda\mathrm{Ai}(\sigma)\mathrm{Ai}(\sigma') + 
\sqrt[3]{\frac{\hbar^2\pi^3}{2m^2g}} [ \Theta(x-x')\mathrm{Ai}(\sigma)\mathrm{Bi}(\sigma') + \Theta(x'-x) \mathrm{Ai}(\sigma')\mathrm{Bi}(\sigma) ] \right\}
\,\,\,.\,\,\,\,\,\,\,\,\,\,\,\,
\end{eqnarray}
\subsubsection{Lloyd interferometer}
We can now take this general result and apply it to the situation in a Lloyd interferometer. For this, we again use the method of mirror charges in order to determine the Green's function:
\begin{eqnarray}\label{Eq:GreensFuncGrav}
G(\mathbf{r},\mathbf{r'}) &=& 4\pi \int \frac{d^2k_\perp}{(2\pi)^2}
\sum\limits_{a,b=\pm} ab\, e^{ \mathrm{i}k_y(y-ay') + \mathrm{i}k_z(z-bz') }
\nonumber
\\
&&
\times
\left\{ \lambda\mathrm{Ai}(\sigma)\mathrm{Ai}(\sigma') + 
\sqrt[3]{\frac{\hbar^2\pi^3}{2m^2g}} [ \Theta(x-x')\mathrm{Ai}(\sigma)\mathrm{Bi}(\sigma') + \Theta(x'-x) \mathrm{Ai}(\sigma')\mathrm{Bi}(\sigma) ] \right\}
\,\,\,.\,\,\,\,\,\,\,\,\,\,\,\,
\end{eqnarray}
Again this Green's function vanishes along the dotted integration path depicted in Fig.\,\ref{Fig:LloydGreen}, and the wave function $\psi(\mathbf{r})$ is only non-vanishing and equals $C e^{\mathrm{i}kz}$ at the entrance slit. Eq.\,(\ref{Eq:PsiLloyd}) still holds. Substituting Eq.\,(\ref{Eq:GreensFuncGrav}) into Eq.\,(\ref{Eq:PsiLloyd}) and taking the $z'$-derivative, we find
\begin{eqnarray}
\psi(\mathbf{r}) &=& -4C\int_{-\infty}^\infty dx' \int^{y_{\mathrm{Sl}} +\delta/2 }_{y_{\mathrm{Sl}} - \delta/2 }  dy' \int \frac{d^2k_\perp}{(2\pi)^2} k_z \sin (k_y y') e^{ \mathrm{i} \mathbf{k}_\perp \cdot \mathbf{x}_\perp }
\nonumber
\\
&&
\times
\left\{ \lambda\mathrm{Ai}(\sigma)\mathrm{Ai}(\sigma') + 
\sqrt[3]{\frac{\hbar^2\pi^3}{2m^2g}} [ \Theta(x-x')\mathrm{Ai}(\sigma)\mathrm{Bi}(\sigma') + \Theta(x'-x) \mathrm{Ai}(\sigma')\mathrm{Bi}(\sigma) ] \right\}
\,\,\,.\,\,\,\,\,\,\,\,\,\,\,\,
\end{eqnarray}
Also evaluating the $y'$-integral and redefining the constant $\sqrt[3]{\frac{2m^2g}{\hbar^2}}\lambda\to\lambda$, we are left with
\begin{eqnarray}
\psi(\mathbf{r}) &=& -8C  \int \frac{d^2k_\perp}{(2\pi)^2} \frac{k_z}{k_y} \sin (k_y y_{\mathrm{Sl}})\sin (k_y \delta/2) e^{ \mathrm{i} \mathbf{k}_\perp \cdot \mathbf{x}_\perp }
\nonumber
\\
&&
\times
\int_{-\infty}^\infty d\sigma'
\left\{ \lambda\mathrm{Ai}(\sigma)\mathrm{Ai}(\sigma') + 
\pi [ \Theta(\sigma-\sigma')\mathrm{Ai}(\sigma)\mathrm{Bi}(\sigma') + \Theta(\sigma'-\sigma) \mathrm{Ai}(\sigma')\mathrm{Bi}(\sigma) ] \right\}
\,\,\,.\,\,\,\,\,\,\,\,\,\,\,\,
\end{eqnarray}
Since \cite{NIST}
\begin{eqnarray}
\int_{-\infty}^\infty d\sigma' \mathrm{Ai}(\sigma') &=& 1\,\,\,,
\end{eqnarray}
and
\begin{eqnarray}
&&\!\!\!\!\!\!\!\!\!\!\!\!\!\!\!\!\int_{-\infty}^\infty d\sigma' [ \Theta(\sigma-\sigma')\mathrm{Ai}(\sigma)\mathrm{Bi}(\sigma') 
+ \Theta(\sigma'-\sigma) \mathrm{Ai}(\sigma')\mathrm{Bi}(\sigma) ]
\nonumber
\\
&&
= \frac{1}{3}\mathrm{Bi}(\sigma) + \frac{\sigma^2}{2\pi}\left[ {}_0F_1\bigg( ;\frac{2}{3};\frac{\sigma^3}{9} \bigg) {}_1F_2\bigg(\frac{2}{3} ;\frac{4}{3},\frac{5}{3};\frac{\sigma^3}{9} \bigg) - 
2
{}_0F_1\bigg( ;\frac{4}{3};\frac{\sigma^3}{9} \bigg) {}_1F_2\bigg(\frac{1}{3} ;\frac{2}{3},\frac{4}{3};\frac{\sigma^3}{9} \bigg)
\right]
\,\,\,,
\end{eqnarray}
we finally obtain
\begin{eqnarray}
\psi(\mathbf{r}) &=& -8C  \int \frac{d^2k_\perp}{(2\pi)^2} \frac{k_z}{k_y} \sin (k_y y_{\mathrm{Sl}})\sin (k_y \delta/2) e^{ \mathrm{i} \mathbf{k}_\perp \cdot \mathbf{x}_\perp }
\nonumber
\\
&&
\times
\left\{ \lambda\mathrm{Ai}(\sigma) + \frac{\pi}{3}\mathrm{Bi}(\sigma)
+
\frac{\sigma^2}{2}\bigg[ {}_0F_1\bigg( ;\frac{2}{3};\frac{\sigma^3}{9} \bigg) {}_1F_2\bigg(\frac{2}{3} ;\frac{4}{3},\frac{5}{3};\frac{\sigma^3}{9} \bigg) \right.
\nonumber
\\
&&
\phantom{\times
\left\{ \lambda\mathrm{Ai}(\sigma) + \frac{\pi}{3}\mathrm{Bi}(\sigma)
+
[\right.}\,\,\,\,\,
\left.
-
2
{}_0F_1\bigg( ;\frac{4}{3};\frac{\sigma^3}{9} \bigg) {}_1F_2\bigg(\frac{1}{3} ;\frac{2}{3},\frac{4}{3};\frac{\sigma^3}{9} \bigg)
\bigg] \right\}
\,\,\,.\,\,\,\,\,\,\,\,\,\,\,\,
\end{eqnarray}


\section{Conclusions}

Neutron Lloyd interferometry is a promising experimental setup that can be used to probe effects within and beyond the realms of known physics. However, previous theoretical analyses made use of approximative methods, which are expected to not always be sufficiently accurate. For this reason, in this article, we presented an exact full Green's functions analysis of Lloyd interferometry. First, we discussed the hypothetical case of no external fields acting on the neutrons while travelling within the experimental setup. For this, we found the corresponding Green's function for solving the Schr\"odinger equation and subsequently determined the resulting wave function. Second, we looked at the physically relevant case of having an external gravitational field acting on the neutrons.
\\
The prediction made for the gravitationally modified neutron wave function serves as a theoretical basis for using a neutron Lloyd interferometer to probe gravitational fields. However, it should be stressed that the computations presented here approximate the interferometer's slit source as being infinitely wide in $x$-direction and the mirror to be infinitely extended. For practically applying the technology developed in this article to a real experiment, a numerical analysis without these approximations would be required. Such a sophisticated numerical evaluation will be subject of future work. Furthermore, in the future, the computation shown in the present article, will serve as a blueprint for predicting neutron wave functions in Lloyd interferometers under the influence of hypothetical gravity-like fifth forces and scalar fields.


\begin{acknowledgments}

The authors thank H.\,Filter and T.\,Jenke for useful discussions. Y.N. Pokotilovski and P. Geltenbort have drawn our attention to this topic as a tool for searches for hypothetical gravity-like interactions.
This work was supported by the Austrian Science Fund (FWF): P 34240-N and P 33279-N. The
authors acknowledge TU Wien Bibliothek for financial support through its Open Access Funding Programme.

\end{acknowledgments}


\bibliography{Lloyd}
\bibliographystyle{JHEP}

\end{document}